\documentstyle[aps,prb,epsfig,multicol]{revtex}

\begin{document}

\title{Nonlocal Effects on the Surface Resistance of High Temperature
Superconductors with (100) and (110) Surfaces}

\author{C. T. Rieck, D. Straub, and K. Scharnberg}
\address{Fachbereich Physik, Universit\"at Hamburg,
Jungiusstr. 9, D-20355 Hamburg, Germany}

\date{\today}
\maketitle

\begin{abstract}
The low temperature surface resistance $R_{\text{s}}$ of $d$-wave
superconductors is calculated as function of frequency assuming normal
state quasiparticle mean free paths $\ell$ in excess of the
penetration depth. Results depend strongly on the geometric
configuration. In the clean limit, two contributions to $R_{\text{s}}$
with different temperature dependencies are identified: photon
absorption by quasiparticles and pair breaking. The size of nonlocal
corrections, which can be positive or negative depending on frequency
decreases for given $\ell$ as the scattering phase shift
$\delta_{\text{N}}$ is increased. However, except in the unitarity
limit $\delta_{\text{N}}=0.5\,\pi$, nonlocal effects should be
observable.
\end{abstract}


\begin{multicols}{2}

\section{INTRODUCTION}
\label{sec:intro}

With the onset of superconductivity, a new length scale, the BCS
coherence length $\xi_{_0} = v_{\text{F}} / \pi\Delta_0$ enters.
\cite{Tinkham}
Its size, rather than the size of the normal state
quasiparticle mean free path $\ell$, relative to the penetration depth
$\lambda$ determines whether or not the electromagnetic response of the
condensate is local.
The situation is different for thermally exited quasiparticles which
are responsible for losses at frequencies below the pair breaking
threshold. These quasiparticles behave very much like quasiparticles
in the normal state once the change in the density of states and the
resulting change in thermal occupation of quasiparticle states are
taken into account.\cite{SCH78}
Hence one must expect to see nonlocal effects in the
surface resistance of superconductors when $\ell \ge \lambda$.
At finite frequencies the quasiparticles contribute to the screening
currents so that there are some nonlocal corrections to the
penetration depth of microwave fields as well.

High-$T_{\text{c}}$ materials are highly anisotropic. They can be
viewed as stacks of weakly coupled conducting planes. In this paper we
shall only consider the in-plane conductivity because it is only this
component of the conductivity tensor which can be expected to show
nonlocal effects. The in-plane coherence length typically is between
10\,\AA\ and 20\,\AA\ while the in-plane penetration depth is two
orders of magnitude larger, so that high-$T_{\text{c}}$
superconductors are well into the London limit \cite{Tinkham},
$\xi_{_0} \ll \lambda$.  While the quasiparticle in-plane mean free
path $\ell$ is comparable to
$\xi_{_0}$ near the transition temperature $T_{\text{c}}$, so that the
corresponding scattering rate $\Gamma = v_{\text{F}}/2\ell$ is of the
order of $T_{\text{c}}$, it is now generally accepted that $\ell$
increases substantially below
$T_{\text{c}}$. \cite{Bonna,Bonnb,Hensen97,Doettinger,Harris} In high 
quality single crystals, the disorder induced mean free path
$\ell^{\text{el}}$ can exceed the London penetration depth by quite a
wide margin.
\cite{Bonna,Bonnb,Hensen97,Rieck}
In contrast to the normal state where a single parameter suffices to
characterize the elastic scattering, both the concentration of
scattering centers and the strength of the individual scatterer or,
alternatively, the normal state elastic scattering rate $\Gamma$ and the
scattering phase shifts have to be specified in the case of
superconductors with unconventional\cite{Louis} order parameters. We
shall show below for the case of $s$-wave scattering that the
scattering phase shift $\delta_{\text{N}}$ plays a very important role.

It has been pointed out by Chang and Scalapino \cite{Scalb} that
nonlocal effects are absent when the conducting planes are parallel to
the surface exposed to the microwave radiation ($\hat{c}$-axis
orientation) because the spatial variation of the fields parallel to
the surface occurs on length scales much larger than the penetration
depth. Thus, for strictly 2\,D quasiparticle motion, the scalar
product ${\bbox{v}}_{\text{F}}\cdot{\bbox{q}}$ vanishes.

Zuccaro \emph{et al.}\cite{Zuccaro} argued that, at least for
YBa$_2$Cu$_3$O$_{7-\delta}$ (YBCO), coherent coupling between planes
is not negligible so that there exists a finite component
${\bbox{v}}_{\text{F}}^{\text{c}}$ of the Fermi velocity parallel to
the surface normal.  Assuming an isotropic gap, these authors
calculated the nonlocal corrections to the surface impedance for
parameters such that $\xi_{_0} \ll \lambda$.  These calculations show
the expected increase in the surface resistance in the anomalous skin
effect regime 
$\omega \le v_{\text{F}}/\lambda =
(\xi_{_0}/\lambda)\pi\Delta_0 \ll \Delta_0$ 
resulting from the direct absorption of photons by quasiparticles in
an energy- and momentum-conserving process.  
Li \emph{et al.}\cite{Hirsch98a} also invoked finite dispersion
perpendicular to the planes to argue that nonlocal effects prevent the
observation of the nonlinear Mei\ss{}ner effect predicted by Yip and
Sauls. \cite{Yip} While this assumption of a 3\,D tight-binding Fermi
surface appears quite reasonable for YBCO, one nevertheless expects
$\bbox{v}_{\text{F}}^{\text{c}} \ll \bbox{v}_{\text{F}}^{\text{ab}}$.
Nonlocal effects will, therefore, be less significant for
$\hat{c}$-axis oriented samples than for samples whose surfaces
contain the $\hat{c}$-axis, like $\hat{a}$-axis oriented films or thin
needles of single crystals with the $\hat{c}$-axis being the longest
dimensions. Setting aside the problem of preparing high-quality
$\hat{a}$-axis oriented films, microwave measurements of the surface
impedance of such films are not well suited to a search for nonlocal
effects because the typical current and field distributions
\cite{Hensen97} involve two vastly different components of the
conductivity tensor. Transmission experiments in the THz regime, using
linearly polarized light, would be more feasible.\cite{Pimenov} For
this reason we investigate the surface impedance for a rather broad
range of frequencies.  At very high frequencies, though, inelastic
scattering would reduce the quasiparticle mean free paths even at low
temperatures to such an extent that the local limit applies.  An
investigation of the electromagnetic response of a needle shaped
single crystal to a parallel microwave magnetic field, which would
involve the in-plane conductivity only, appears to be even more
promising in view of the thick single crystals that can now be grown
in BaZrO$_3$ crucibles.
\cite{Erb}

If the pairing state in high-$T_{\text{c}}$ superconductors has nodes, which
is at present a widely held belief, then a momentum dependent coherence
length could be introduced. This would exceed the penetration  depth
for momentum states in the nodal region so that  high-$T_{\text{c}}$ materials
are London superconductors only in the sense that for the majority of
$\bbox{k}$-states $\xi_{_0} (\bbox{k}) \ll \lambda$. This point was first
raised by Kosztin and Leggett, \cite{Leggett}
who found that the zero frequency
clean limit penetration depth varies quadratically with temperature at
very low temperatures, rather than linear which is considered to be a
hallmark of $d$-wave superconductivity. 
This change in the temperature dependence of $\lambda$ due to
nonlocality is however easily masked by mean free path and finite
frequency effects.\cite{Hensen97}

Schopohl and Dolgov\cite{SD1} have argued that a linear temperature
dependence of the penetration depth would violate the third law of
thermodynamics. The nonlocal effects discovered by Kosztin and Leggett
\cite{Leggett}
could reconcile the $d$-wave model with this general thermodynamic
argument but only for the geometry in which the $\hat{c}$-axis is parallel
to the sample surface. Hirschfeld \emph{et al.} \cite{Hirsch98b}
pointed out that the electromagnetic response kernel contains, in
addition to the
$\bbox{v}_{\text{F}} \cdot \bbox{q}$ term, a term $q^2/2m$ which
leads to nonlocal corrections at some very low temperatures in any
geometry. For $q \approx 1/\lambda \ll k_{\text{F}}$ this term can
usually be neglected and this is what we shall do here because
we are interested in microwave losses at low but finite
temperatures. For strictly 2\,D systems, one
might wonder whether fluctuation effects would not be much more
important than such very small nonlocal corrections.

Another consequence of $d$-wave pairing is the fact that pair-breaking
is possible at any frequency. However, for this process to contribute
to the transverse conductivity, the finite photon momentum needs to be
taken into account unless the required momentum transfer is provided
by some other scattering event occuring simultaneously. Since we are
interested here in low temperatures, we shall only consider disorder
induced elastic scattering. If only $s$-wave scattering is taken into
account, it is easy enough to write down a general expression for the
complex conductivity $\sigma(\bbox{q}, \omega)$. In order to isolate
the two contributions to the surface resistance which are due to
nonlocality, we shall also consider the clean limit.

\section{THEORY, GENERAL CASE}
\label{sec:theory,general}

The current-current correlation function
from which the transverse conductivity is derived according to
\begin{equation}
\sigma({\bbox{q}}, \omega) = - \frac{e^2}{i\omega}\,
\left\{
  \frac{n}{m} + \big\langle[j_y,j_y]\big\rangle(\bbox{q},\omega )
\right\}\,,
\label{eqn:ta1}
\end{equation}
can be expressed very simply in terms of normal and anomalous single
particle Green's functions \cite{Rick}
\begin{eqnarray}
\lefteqn{
\big\langle[j_y,j_y] \big\rangle(\bbox{q},i\nu_m ) =
2T \sum_{\omega_n} \int\frac{d^Dp}{(2\pi)^D} \, \frac{p_y^2}{m^2}\, 
\times
}\;
\nonumber\\
&&
\times
\Big[
   {\cal G}_{\omega_n}(\bbox{p})\, 
   {\cal G}_{\omega_n - \nu_m}(\bbox{p}\!-\!\bbox{q})
  +{\cal F}_{\omega_n}(\bbox{p})\,
   {\cal F}_{\omega_n - \nu_m}(\bbox{p}\!-\!\bbox{q})
\Big]
\, ,
\label{eqn:ta2}
\end{eqnarray}
when there are no vertex corrections. This is the case for isotropic
disorder induced scattering, which we will focus on here. Momentum
dependent inelastic interactions, capable of causing the formation of
unconventional superconducting pair states, do require consideration
of vertex corrections. \cite{Pinesb} At low temperatures and low
frequencies, however, the contributions of these interactions to the
quasiparticle lifetimes are negligible. All the complications
resulting from quasiparticles in an unconventional superconductor
being scattered off point defects, which were first discussed in the
context of Heavy Fermion superconductors, \cite{Louis} affect only the
single particle selfenergies discussed below.

The momentum integral in Eq.\ (\ref{eqn:ta2})
is evaluated  under the assumption that the main contribution
comes from quasiparticle states near the Fermi surface.
We consider purely two-dimensional conduction and simplify the 2D
Fermi surface to a circle. The Fermi velocity $v_{\text{F}}$
and the density of states per spin at the Fermi level
$N(0) = m/2\pi $ are combined into a single parameter,
the plasma wavelength $\lambda_{\text{p}}$, according to
\begin{equation}
\frac{1}{\mu_{_0}\lambda_{\text{p}}^2} 
= e^2\,N(0)\,v_{\text{F}}^2 \frac{1}{s}
\, ,
\label{eqn:ta3}
\end{equation}
where $s$ is the average distance between conducting planes.
In the clean limit, $\lambda_{\text{p}}$  is identical to the zero
temperature London penetration depth $\lambda_{\text{L}}$. With these
assumptions one obtains for the conductivity after analytic
continuation of $i\nu_m$ to the real axis
\begin{eqnarray}
\lefteqn{
\sigma (\bbox{q},\omega ) =
\frac{1}{\mu_{_0}\lambda_{\text{p}}^2} \, \frac{1}{2\omega} \,
\int_{-\omega /2}^{+\infty} d\Omega
\int_{0}^{2\pi} \frac{d\varphi}{2\pi}\,	\sin^2\varphi
} \;\;
\nonumber\\
&&
\bigg\{\;\;\,
\tanh \frac{\Omega +\omega}{2T} \,
2i\,{\rm Im}\,
M(\varphi ; \bbox{q} ;\Omega_+ +\omega ,\Omega_+ )
\nonumber\\
& &
\nonumber\\
& &
\;
-\left[
    \tanh \frac{\Omega\!+\!\omega}{2T} -\tanh \frac{\Omega}{2T} 
\right]\,
    M(\varphi ; \bbox{q} ; \Omega_+ +\omega ,\Omega_+ )
\nonumber\\
& &
\nonumber\\
& &
\;
+\left[
   \tanh \frac{\Omega\!+\!\omega}{2T} - \tanh \frac{\Omega}{2T} 
\right]\,
M(\varphi ; \bbox{q} ; \Omega_+ +\omega ,\Omega_- )
\bigg\}
\label{eqn:ta4}
\end{eqnarray}
where $+(-)$ indicates a positive (negative) infinitesimal imaginary part.
$M$ is an abbreviation for
\begin{eqnarray}
\lefteqn{
M(\varphi ; \bbox{q} ; \Omega_+ + \omega ,\Omega_\pm ) =
} 
&&
\nonumber\\
&&
\nonumber\\
&&
\left.\left(1 + 
   \frac{(\Omega  + \omega ) \, Z(\Omega _+ + \omega ) \, 
         \Omega \, Z(\Omega _\pm ) + \Delta^2(\varphi )}%
        {R_1\,R_2}
\right.\right)
\times
\nonumber\\
& &
\nonumber\\
& &
\times 
\frac{R_1 + R_2}%
     {\big\lbrack R_1 + R_2 \big\rbrack^2
      + 
      \big\lbrack\, v_{\text{F}} q\cos\varphi
        + \chi (\Omega _\pm )-\chi (\Omega _+ + \omega )
      \big\rbrack^2}
\label{eqn:ta5}
\end{eqnarray}
where
\begin{equation}
\renewcommand{\arraystretch}{2}
\begin{array}{rcl}
R_1 &=& \displaystyle 
      \sqrt{ \Delta^2(\varphi ) -
      \big(  \Omega\,           Z(\Omega _\pm       )\big)^2}\,,
\\
R_2 &=& \displaystyle 
      \sqrt{ \Delta^2(\varphi ) -
      \big( (\Omega + \omega)\, Z(\Omega _+ + \omega)\big)^2}\,.
\label{eqn:ta6}
\end{array}
\end{equation}
The order parameter, which is assumed to have d$_{x^2-y^2}$-symmetry
with the respect to the crystallographic axes, is represented as
\begin{equation}
\Delta(\varphi) = \Delta_0(T)\,\cos 2(\varphi\!-\!\varphi_{_0})\,,
\label{eqn:ta7}
\end{equation}
where $\varphi_{_0}$ is the angle between the crystallographic
(100)-axis and the surface normal. In the local limit, the orientation
of the order parameter relative to the sample surface has no effect on
the conductivity, provided one neglects the suppression of the order
parameter by the surface\cite{Buchholtzb}
which occurs on a length scale 
$\xi_{_0} \ll \lambda_{\text{p}}$. 
The $\varphi$-integral can then be reduced to the
interval $[0,\pi/2]$. In the nonlocal case with $\varphi_{_0}$
arbitrary such a reduction produces four different terms.  For the
exceptional cases $\varphi_{_0} = 0$ [(100)-oriented surface] and
$\varphi_{_0} = \pi/4$ [(110)-oriented surface], these terms are
pairwise equal. Further simplification is possible in the clean limit
(see next section), as well as in the strong and weak scattering
limit, because in these limits the selfenergy $\chi$ either vanishes
or becomes independent of $\Omega$, so that $M$ depends only on
$\left\lbrack\, v_{\text{F}} q\cos\varphi\right\rbrack^2$.

The dependence of $M$ on $\bbox{q}$ is very  simple because we neglected
$\bbox{q}$ in the argument of the order parameter  and the other
self-energies. Anticipating $\vert\bbox{q}\vert \approx
1/\lambda_{\text{p}}$ this is justified.
If one wanted to keep, in addition to $v_{\text{F}} q\cos\varphi $,
the term $q^2/2m$,\cite{Hirsch98b} such an approximation would be
inconsistent. For $q \approx k_{\text{F}}$ the response of a $d$-wave
superconductor changes its character completely, because the coherence
factors would revert from case~II to case~I.\cite{Tinkham,Flatte}

Because of the restriction to $s$-wave scattering there are
no self-energy corrections to the $d$-wave order parameter.
The remaining selfenergy corrections are to be determined from
\begin{equation}
\Omega	Z(\Omega _\pm ) =
\Omega	+ \Gamma_{\text{N}}^{\text{el}}
\frac{\langle  g_{_0}(\varphi  ,\Omega _\pm\!) \rangle}%
     {\cos^2 \delta_{\text{N}} \!-\! \sin^2 \delta_{\text{N}}
      \langle g_{_0}(\varphi  ,\Omega _\pm\!) \rangle^2 }
\label{eqn:ta8}
\end{equation}
and
\begin{equation}
\chi (\Omega _\pm ) =
\Gamma_{\text{N}}^{\text{el}} \,
\frac{\cot \delta_{\text{N}}}%
     {\cos^2 \delta_{\text{N}} - \sin^2 \delta_{\text{N}}
      \langle  g_{_0}(\varphi  ,\Omega _\pm\!)  \rangle^2 }
\label{eqn:ta9}
\end{equation}
$\langle g_{_0} \rangle$ is the energy-integrated normal Green's
functions, averaged over the Fermi circle
\begin{equation}
\langle g_{_0} (\varphi ,\Omega _\pm ) \rangle =
 \int\limits_0^{2\pi} \frac{d\varphi}{2\pi}
 \frac{\Omega  Z(\Omega _\pm )}%
      {\sqrt{ \Delta^2(\varphi ) - \left(\Omega Z(\Omega_\pm ) \right)^2}}
\label{eqn:ta10}
\end{equation}

\begin{equation}
\Gamma_{\text{N}}^{\text{el}}
 = n_{\text{imp}}\,\frac{\pi N(0)v^2}{1 + \left(\pi N(0) v\right)^2}
 = \frac{n_{\text{imp}}}{\pi N(0)}\,\sin^2\delta_{\text{N}}
\label{eqn:ta11}
\end{equation}
is the elastic scattering rate in the normal state
and $\delta_{\text{N}} = \tan^{-1}(\pi N(0) v)$ is the scattering 
phase shift.

From $\sigma(\bbox{q}= q \hat{e}_z\,,\,\omega )$ the
surface impedance is calculated assuming specular reflection
\begin{eqnarray}
{\cal Z}_{\text{s}} 
&=&
R_{\text{s}}(\omega ) - i\omega \mu_{_0}\lambda (\omega )
\nonumber\\
&=&
  - i\omega \mu_{_0}\, \frac{2}{\pi}\,\int\limits_0^\infty 
  \frac{dq}{q^2 - i\omega \mu_{_0} \sigma (q,\omega )}
\; .
\label{eqn:ta12}
\end{eqnarray}
We have not evaluated the formula applicable for diffuse surface
scattering for lack of computer time.

\section{THEORY, CLEAN LIMIT}
\label{sec:theory,clean}

Since we are primarily interested in the surface resistance, we shall
calculate only the real part $\sigma_1$ of the complex conductivity
Eq.~(\ref{eqn:ta1}).  The imaginary part $\sigma_2$ should simply be
given by\cite{Tinkham,Hensen97} $1/\omega\mu_{_0}\lambda_{\text{p}}^2$
and this expectation is borne out by the numerical calculations based
on the general formalism presented in the previous section.  Inserting
spectral representations for the Green's functions, evaluating the sum
over Matsubara frequencies $\omega_n$ and performing the analytic
continuation with respect to $i\nu_m$ we obtain
\begin{eqnarray}
\label{eqn:tb1}
\lefteqn{
\sigma_1 (\bbox{q},\omega ) =
\frac{1}{\mu_{_0}\lambda_{\text{p}}^2}\, \frac{1}{\omega }\,
 \int\limits_{-\omega/2}^\infty d\Omega
		\big[ f(\Omega) - f(\Omega+\omega)\big]
\times
}\qquad\;
\nonumber\\
\times
&&
  \int\limits_0^{2\pi} d\varphi \,\sin^2\varphi\,
     I_j \Big[
	   \Omega,\Omega+\omega,v_{\text{F}} q\cos\varphi,\Delta(\varphi)
	 \Big]
\end{eqnarray}
with
\begin{eqnarray}
\label{eqn:tb2}
I_j &=& 
\frac{\pi}{4}\int\limits_{-\infty}^\infty \frac{d\xi}{2\pi}
      \Big[
	\mbox{Im}\, G_{\Omega_+}(\xi) \,
	\mbox{Im}\, G_{\Omega_+ - \omega}(\xi-v_{\text{F}} q
	\cos\varphi)
\nonumber\\
&&
      {}+ \mbox{Im}\, F_{\Omega_+}(\xi) \,
	  \mbox{Im}\, F_{\Omega_+ - \omega}(\xi-v_{\text{F}} q \cos\varphi)
      \Big]\,.
\end{eqnarray}
Since
\begin{eqnarray}
\label{eqn:tb3}
\lefteqn{
\mbox{Im}\, G_{\Omega_+}(\xi) = -\frac{\pi}{2} 
\times
}\qquad
\nonumber\\
&&
  \sum_{\sigma=\pm1}
  \left( 
    \sigma + \frac{\Omega}{\sqrt{\Omega^2 - \Delta^2}} 
  \right)
  \delta\left( \xi - \sigma\sqrt{\Omega^2 - \Delta^2} \right)
\end{eqnarray}
the energy integral is easily done, yielding
\begin{eqnarray}
\label{eqn:tb4}
\lefteqn{
I_j = \frac{\Omega}{|\Omega|}
\times
}
\nonumber\\
&&
\;
\sum_{\sigma=\pm 1}
\left(
   \sigma + \frac{\Omega(\Omega+\omega) + \Delta^2}%
                 {\sqrt{\Omega^2 - \Delta^2}
		  \sqrt{(\Omega+\omega)^2 - \Delta^2}}
\right)
\delta \big[ g_\sigma (\Omega) \big]
\end{eqnarray}
where
\begin{eqnarray}
\label{eqn:tb5}
g_\sigma (\Omega) 
&=&   
v_{\text{F}} q\cos\varphi
  - \sqrt{(\Omega+\omega)^2 - \Delta^2(\varphi)}
\nonumber\\
&&
{}+ \sigma \sqrt{\Omega^2 - \Delta^2(\varphi)}
\end{eqnarray}

So far, the evaluation of $\sigma_1$ is equivalent to that given by
Mattis and Bardeen\cite{Mattis} who, at this point, take the extreme
anomalous limit, i.e.\ they neglect the $\Omega$-dependence of
$g_\sigma $.
If this approximation were used in the case of a $d$-wave
superconductor with nodal lines parallel to the surface, the normal
state conductivity
$\sigma_1 = \frac{3\pi}{4} \,\frac{1}{\mu_{_0}\lambda_{\text{p}}^2}\,
\frac{1}{v_{\text{F}} q}$,
appropriate for the anomalous skin effect regime, would be obtained.
Since HTC materials are essentially London superconductors, taking
this limit is not justifiable. Instead, we shall find the zeros of
$g_\sigma (\Omega)$ and perform the $\Omega$-integral. In this way we
cannot reproduce the normal state result because, for $\Delta \to 0$,
$g_\sigma$ becomes either independent of $\Omega$ or the
$\delta$-functions give a vanishing contribution.

However, the same approach is used successfully in the theory of the
electronic Raman response of high temperature superconductors, which
involves a density-density correlation function with some vertex
$\gamma(\varphi )$
specific to the Raman response in place of the current vertex $p_y/m$.
The important difference is that the coherence factor is
case~I,\cite{Tinkham} i.e.\ $\Delta^2$ in the numerator of Eqs.\
(\ref{eqn:ta5},\ref{eqn:tb4}) has a different sign. One can then
take the local limit $q \to 0$ and evaluate the frequency integral
using $g_{-1}(-\omega/2) = 0$ with the result
\begin{eqnarray}
\lefteqn{
-\,{\rm Im}\, \big\langle [n,n] \big\rangle\left(0, \omega \right) 
=
\frac{2 N(0)}{\omega}\, \tanh \left(\frac{\omega}{4T} \right)
\times
}
\nonumber\\
&&
\quad
\times
\int\limits_0  ^{\pi/2} d \varphi \,
\gamma^2 \left(\varphi \right) \big[ 2 \Delta
\left( \varphi, T \right)\big]^2
\frac{\theta\,\big[\, \omega\!-\!\vert 2\Delta\!
\left( \varphi, T \right)\!\vert \big]}
{\sqrt{\omega^2\!-\!\left[2 \Delta\!\left( \varphi, T \right)
\right]^2}}
\,.
\label{clean}
\end{eqnarray}
This pair breaking contribution to the Raman response does account
quite successfully  \cite{Dev95b,Einz96} for a range of experimental
observations.
The corresponding result for $\sigma_1$ is, of course, zero.
This difference in the local clean limit explains why the
{\it absorption} of low frequency photons by charge carriers in high
temperature superconductors depends much more strongly on
quasiparticle mean free paths
than the {\it inelastic scattering} of such photons.
\cite{Krim}

Solving $g_{\pm 1}(\Omega) = 0$ in the interval $[-\omega/2, \infty]$
leads to
\begin{equation}
\label{eqn:tb6}
\Omega_0(\varphi) =
  - \frac{\omega}{2}
  + \frac{1}{2}  v_{\text{F}} q\cos\varphi
    \sqrt{1 - \frac{[2\Delta(\varphi)]^2}%
		   {\omega^2 - [v_{\text{F}} q\cos\varphi]^2}}
\; .
\end{equation}
For $g_{+1[-1]}\big( \Omega_0(\varphi) \big)$
to vanish,
$\omega
\sqrt{1 - \frac{[2\Delta(\varphi)]^2}%
	       {\omega^2 - [v_{\text{F}} q\cos\varphi]^2}}
$
must be greater [less] than $v_{\text{F}} q\cos\varphi$ so that
either $\delta \left[ g_{+1}\right]$ or $\delta \left[ g_{-1}\right]$
contributes.
$\Omega_0(\varphi)$ must be real, which requires
\begin{equation}
\label{eqn:tb7}
1 - \frac{[2\Delta(\varphi)]^2}%
	 {\omega^2 - [v_{\text{F}} q\cos\varphi]^2} \ge 0
\; .
\end{equation}
For $\cos \varphi > \frac{\omega}{v_{\text{F}} q}$ this is always
fulfilled, irrespective of the magnitude and $\varphi$-dependence of
the order parameter. $\Omega_0(\varphi)$ is positive for such
$\varphi$ so that this represents the quasiparticle contribution to
$\sigma_1$ which, because of the Fermi function in (\ref{eqn:tb1}),
becomes small at low temperatures.  In the opposite case
$\Omega_0(\varphi)$ is negative. The inequality (\ref{eqn:tb7}) then
leads to the usual condition $\omega > 2\Delta$ for pair breaking,
when the order parameter is isotropic.
For a $d$-wave superconductor, $\Delta(\varphi)$ has to be specified
at this point. We introduce the abbreviations
\begin{equation}
t \equiv \cos \varphi, \quad
y \equiv \left[ \frac{2\Delta_0(T)}{\omega}  \right]^2, \quad
x \equiv \left[ \frac{v_{\text{F}} q}{\omega}  \right]^2
\; .
\end{equation}
For a (100)-surface we can then write
\begin{equation}
\Delta(\varphi) = \Delta_0(T) \left[ 2t^2 - 1  \right]
\; .
\end{equation}

For this orientation of the $d$-wave order parameter
the inequality (\ref{eqn:tb7}) is fulfilled for
$t \in \left[ t_{\text{min}}, t_{\text{max}} \right]$
with
\begin{eqnarray}
\label{eqn:tb8}
\lefteqn{
t_{\text{min}} =
}
\nonumber\\
&&
\max
  \left\{
     0, \left[ \textstyle
	   \frac{1}{2} \left( 1 - \frac{x}{4y} \right)
	  -\frac{1}{2}
	   \sqrt{\left( 1 - \frac{x}{4y} \right)^2 \! + \frac{1}{y} - 1}
	\right]^{\frac{1}{2}}
  \right\}\!
\\
\label{eqn:tb9}
\lefteqn{
t_{\text{max}} =
}
\nonumber\\
&& 
\min
  \left\{
     1, \left[ \textstyle
	   \frac{1}{2} \left( 1 - \frac{x}{4y} \right)
	  +\frac{1}{2}
	   \sqrt{\left( 1 - \frac{x}{4y} \right)^2 \! + \frac{1}{y} - 1}
	\right]^{\frac{1}{2}}
  \right\}
\!.
\end{eqnarray}
Such an interval exists only if the square root is real. For
$y \le 1 \Longleftrightarrow \omega \ge 2\Delta_0$ this is always the
case. For $y > 1$, the interval is empty if $x > 4y$. For $x < 4y$,
the square root is real provided
$x \le 4y \left[ 1 - \sqrt{1 - 1/y} \right]$.
For $\omega \ll 2\Delta_0$ this limits the $q$-integration in
Eq.~(\ref{eqn:ta12}) to the interval $\left[ 0, \sqrt{2}\omega \right]$.

Combining Eqs.~(\ref{eqn:tb1}), (\ref{eqn:tb4}), and 
(\ref{eqn:tb6}) we obtain
\begin{eqnarray}
\label{eqn:tb12}
\lefteqn{
\sigma_1(\bbox{q},\omega)
=
  \frac{2(2\Delta_0(T))^2}{\mu_{_0} \lambda_{\text{p}}^2 \omega}
\times
}
\nonumber\\
&&
\times
  \left[
     \Theta \Big( \sqrt{2}\omega - v_{\text{F}} q \Big)
     \int\limits_{t_{\text{min}}}^{t_{\text{max}}} dt
     +
     \Theta \Big( v_{\text{F}} q - \omega \Big)
     \int\limits_{\frac{\omega}{v_{\text{F}} q}}^1 dt
  \right]
\times
\nonumber\\
&&
  \times
  \sqrt{1-t^2}
  \Big[ 
    f \big(\Omega_0(t) \big) - f \big(\omega + \Omega_0(t) \big)
  \Big]
\times
\nonumber\\
&&
\times
  \frac{(v_{\text{F}} q t)^3 \left[ 2t^2-1 \right]^2}%
       {\big[\omega^2 - (v_{\text{F}} q t)^2\big]^2
	\big[2\Omega_0(t) + \omega\big]}
\end{eqnarray}
where the first integral represents the pair breaking contribution
while the second represents the contribution from quasiparticles.

For a (110)-surface, $\Delta(\varphi) = \Delta_0(T)\,\sin 2\varphi$,
so that the factor $\left[ 2t^2-1 \right]^2$ in (\ref{eqn:tb12}) has
to be replaced by $4 \left( 1 - t^2 \right) t^2$. The limits of
integration in the pair breaking contributions also need to be
changed. Instead of the interval 
$\left[ t_{\text{min}}, t_{\text{max}} \right]$ around $\varphi=\pi/4$
we now have two intervals $t \in \left[ 0, t_{\text{min}}' \right]$ and
$t \in \left[ t_{\text{max}}', 1 \right]$ near $\varphi=0$ and
$\varphi=\pi/2$ which can contribute to $\sigma_1$.
$ t_{\text{min}}'$and $ t_{\text{max}}'$ are given by expressions very
similar to (\ref{eqn:tb8}) and (\ref{eqn:tb9}).

\section{Results and Discussion}

Parameters required as input for the numerical calculations are the
plasma wavelength $\lambda_{\text{p}}$, the in-plane Fermi velocity
$v_{\text{F}}$, the transition temperature $T_{\text{c}}$ and
the order parameter amplitude $\Delta_0(T=0)$. The values chosen are
typical of YBCO:\cite{Hensen97}
$T_{\text{c}}=90.5$\,K, $v_{\text{F}}=1.4\cdot 10^7$\,cm/s,
$\lambda_{\text{p}}=140$\,nm, and $2\Delta_0(T=0)/T_{\text{c}}=7.4$.
Because of (\ref{eqn:ta3}), $v_{\text{F}}$ and $\lambda_{\text{p}}$
are not entirely independent. The values given are consistent with an
effective mass $m \approx 3\,m_0$.
$T_{\text{c}}$ sets the temperature scale but is irrelevant
otherwise. The rather large value of $\Delta_0(T=0)$ is deduced from
the sharp drop in $R_{\text{s}}$ and $\lambda(T)$ observed in the
vicinity of $T_{\text{c}}$. If this were attributable to a non-BCS
temperature dependence of $\Delta_0(T)$, a smaller value of
$\Delta_0(T=0)$ could be inferred which would increase all theoretical
predictions for the low temperature surface resistance.

$\lambda_{\text{p}}$ and $v_{\text{F}}$ are the parameters which
control the importance of nonlocal effects. We have performed some
calculations with $\lambda_{\text{p}} = 100$nm. Since in the local limit
$R_{\text{s}} = 0.5\,\omega^2\mu_{_0}^2\lambda_{\text{p}}^3\sigma_1$
is proportional to $\lambda_{\text{p}}$ one has to scale 
$R_{\text{s}}^{\text{local}}$ with $\lambda_{\text{p}}$ to appreciate
the differences in the nonlocal results. Because the effect of modest
changes  in $\lambda_{\text{p}}$ are rather obvious, we shall not
display these results.

Fig.~\ref{fig:1} shows results for $R_{\text{s}}$ at 
temperatures $T=0.13\,T_{\text{c}}$ and $T=0.05\,T_{\text{c}}$ for a
disorder induced scattering rate $\Gamma=0.1\,$meV, which would
correspond to a normal state mean free path of 460\,nm. The scattering
phase shift has been chosen as $\delta_{\text{N}}=0$ (Born
approximation). Allowing for nonlocality yields peaks in
$R_{\text{s}}$ at around 100\,GHz which greatly exceed the results in
the local limit. The peak heights decrease rapidly with decreasing
temperature, which indicates that these contributions to
$R_{\text{s}}$ are due to direct photon absorption by thermally
excited quasiparticles. This process ceases to be effective when
$\omega \gg v_{\text{F}} q$. With $q \approx 1/\lambda_{\text{p}}$ we
estimate that this ``anomalous skin effect regime'' should end at
frequencies around 
$\nu \ge v_{\text{F}} / 2\pi\lambda_{\text{p}} \approx 160\,$GHz 
in agreement with our numerical calculations.
It is remarkable that here we have a 
range of frequencies in which
$R_{\text{s}}$ is predicted to drop quite rapidly.

\noindent
\begin{minipage}{8.6cm}
\centering
\begin{figure}[htb]
\mbox{\psfig{file=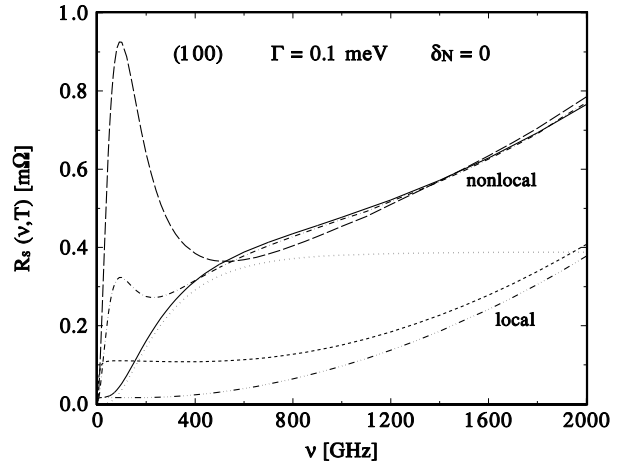,width=8.5cm}}
\caption{\label{fig:1}
    Surface resistance as function of frequency for two temperatures, 
    $T/T_{\text{c}}=0.13$ (dashed lines) and 
    $T/T_{\text{c}}=0.05$ (dot-dashed and dot-dot-dashed lines), 
    for a $d_{x^2-y^2}$ order parameter with its lobes perpendicular 
    and parallel to the surface. 
    The dotted line is the contribution to $R_{\text{s}}$
    from pair breaking in the clean limit. The solid line represents
    the sum of this dotted line and the dot-dot-dashed line.}
\end{figure}
\end{minipage}

\noindent
\begin{minipage}{8.6cm}
  \centering%
  \begin{figure}[htb]
  \mbox{\psfig{file=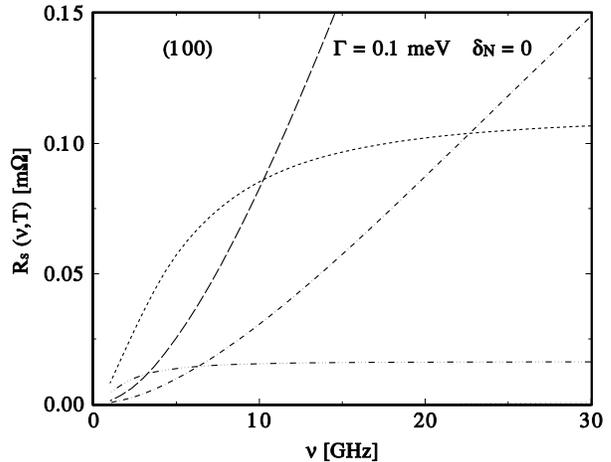,width=8.5cm}}
  \caption{\label{fig:2}
           Blow-up of the low frequency section of
	   Fig.~\ref{fig:1}. Pair breaking effects are negligible at
	   these frequencies.}
  \end{figure}
\end{minipage}

\bigskip

At higher frequencies, $R_{\text{s}}$ increases again but is
essentially temperature independent. This, one would suspect, is the
contribution to $R_{\text{s}}$ from pair breaking. To check this, we
calculated the pair breaking contribution in the clean limit from
(\ref{eqn:tb12}). The result is shown as dotted line. Taking the sum
of this contribution and the result of the local approximation at
$T=0.04\,T_{\text{c}}$ gives the solid line, which reproduces the
result of the full nonlocal calculation remarkably well. The increase
in $R_{\text{s}}$ found in the local approximation must also be
attributed to pair breaking, made possible by the $d$-wave character
of the pair state and the momentum uncertainty due to static disorder.

While Fig.~\ref{fig:1} shows the expected increase in $R_{\text{s}}$,
a completely different picture emerges when the same results are
replotted for the small range of microwave frequencies relevant for
applications (Fig.~\ref{fig:2}). Here we see that, depending on
temperature, there exists a frequency below which $R_{\text{s}}$ in
the local limit {\em exceeds} the results from the nonlocal
calculation. This is a consequence of the particular orientation of
the order parameter we assumed. At these low frequencies the term
$v_{\text{F}} q\cos\varphi$ in (\ref{eqn:ta5}) emphasizes the
contribution from quasiparticles moving parallel to the surface. Since
this is the direction in which the order parameter has its maximum,
the number of occupied $\bbox{k}$-states, that are effective in the
absorption process, is reduced. The change in sign of the nonlocal
correction to $R_{\text{s}}$ with frequency would help to explain why
theoretical predictions based on the local limit are higher than the
low temperature data taken by Bonn {\em et al.}\cite{Bonnb} on
untwinned YBCO single crystals at 4.13\,GHz and lower than those taken
at 34.8\,GHz.\cite{Rieck}

When $R_{\text{s}}$ becomes independent of frequency in the local
limit, $\sigma_1$ must vary as $1/\omega^2$. In the normal state this
would be the case for frequencies such that $\omega/2\Gamma \gg 1$,
for parameters used here $\nu \gg 50\,$GHz. In the superconducting
state, the effective scattering rate, which for given microwave
frequency involves some temperature dependent frequency
average,\cite{Hirschb,Hensen97} depends very sensitively on
$\delta_{\text{N}}$.  For $\delta_{\text{N}} = \pi/2$ it is usually
found to be larger than $\Gamma$, while for $\delta_{\text{N}} = 0$ it
is much less than $\Gamma$ [Ref.~\onlinecite{Hensen97}, Figs.~14 and
19].  Since $\delta_{\text{N}} = 0$ was chosen to obtain the data
shown in Figs.~\ref{fig:1} and \ref{fig:2}, it is not surprising that
$R_{\text{s}}$ becomes frequency independent at frequencies much lower
than the 50\,GHz estimated above.  Since the surface resistance in
$\hat{c}$-axis oriented samples, where we believe the local limit to
apply, does increase between 1\,GHz and 87\,GHz, we must conclude that
the scattering phase shift is closer to $\pi/2$ than 0.

The fact that the full nonlocal results for $R_{\text{s}}$ are lower
at low frequencies than results obtained by taking the local limit
does not depend in any essential way on the scattering rate or the
scattering phase shift. This can be seen from Fig.~\ref{fig:3} where
both types of results, obtained for a range of scattering rates
$\Gamma$ and $\delta_{\text{N}} = 0.4\,\pi$, are compared in a double
logarithmic plot. While $R_{\text{s}}$ depends very sensitively on
$\Gamma$ in the local limit, it is very nearly independent of
scattering when nonlocality is taken into account, as one would expect
in the extreme anomalous limit. Perhaps surprisingly then, and
contrary to the behavior in the normal state, it is
$R_{\text{s}}^{\text{local}}$ that moves towards
$R_{\text{s}}^{\text{nonlocal}}$ to close the gap between the two,
when $\Gamma$ increases. Unlike the high frequency regime shown in
Fig.~\ref{fig:1}, $R_{\text{s}}$ at these microwave frequencies is
certainly not the sum of two independent contributions. The dotted
line is the local result for the lower temperature already shown in
Fig.~\ref{fig:1}. Comparison with the other local curves indicates
that in the superconducting state a small scattering phase shift is
indeed more or less equivalent to a much reduced normal state
scattering rate together with a large scattering phase shift.

\noindent
\begin{minipage}{8.6cm}
  \centering%
  \begin{figure}[htb]
  \mbox{\psfig{file=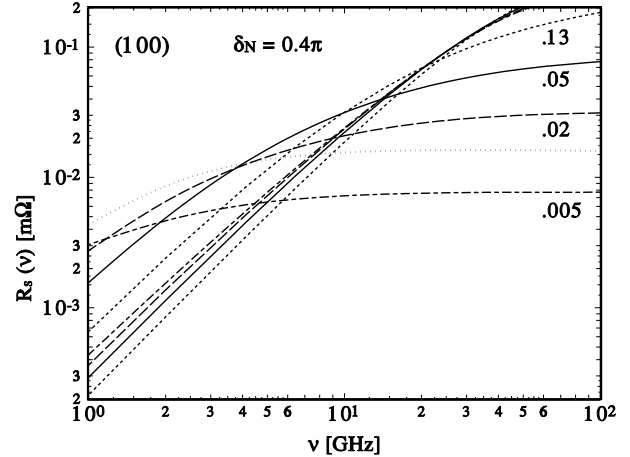,width=8.5cm}}
  \caption{\label{fig:3}
           Double logarithmic plot of the local and nonlocal surface
	   resistance for frequencies 1\,GHz~$\le \nu \le 100\,$GHz
	   and for different scattering rates $\Gamma$. The curves
	   marked with numbers are the results of the local
	   approximations. The numbers themselves are the values of
	   $\Gamma$ in [meV] used in the calculations. The four
	   corresponding results of the fully nonlocal theory are
	   those which attain the lowest values at 1\,GHz. The dotted
	   curve is the local result for $\Gamma=0.1\,$meV and
	   $\delta_{\text{N}}=0$.} 
  \end{figure}
\end{minipage}

\noindent
\begin{minipage}{8.6cm}
  \centering%
  \begin{figure}[htb]
  \mbox{\psfig{file=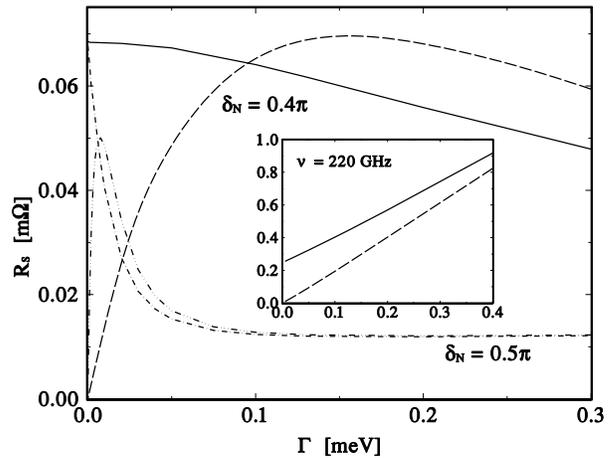,width=8.5cm}}
  \caption{\label{fig:4}
           Surface resistance as function of the scattering rate for
	   two different scattering phase shifts for fixed frequency
	   $\nu=20\,$GHz. The inset shows $R_{\text{s}}$ for
	   $\delta_{\text{N}}=0.4\,\pi$ at $\nu=220\,$GHz. Local
	   results are those that vanish for $\Gamma \to 0$. The
	   temperature is $T/T_{\text{c}}=0.04$.}
  \end{figure}
\end{minipage}

\bigskip

The dependence of $R_{\text{s}}$ on $\Gamma$ and $\delta_{\text{N}}$
at a fixed frequency is further elucidated in the main frame of
Fig.~\ref{fig:4}. In the unitarity limit 
$\delta_{\text{N}} = 0.5\,\pi$, 
nonlocal effects vanish for $\Gamma \ge 0.1\,$meV. In this
case the value of $R_{\text{s}}$ can be calculated from the
``universal'' conductivity 
$\sigma_{00} = 1/\mu_{_0}\lambda_{\text{p}}^2\pi\Delta_0(0)$.\cite{PAL} 
In the local limit, $R_{\text{s}}$ must vanish for $\Gamma \to 0$.
Before it does so, $R_{\text{s}}(\Gamma)$ goes through a 
maximum, whose height, width and position depend sensitively on
$\delta_{\text{N}}$. Because of this behavior one can fit the low
temperature surface resistance of high quality YBCO samples by
choosing a small value of $\Gamma$ which is, however, still compatible
with the expected structural disorder, and then adjusting the
scattering phase shift.\cite{Hensen97} With nonlocality taken into
account, $R_{\text{s}}$ has a finite limit for $\Gamma \to 0$, as in
the normal state. At the lower end of the microwave regime
$R_{\text{s}}^{\text{nonlocal}} \gg R_{\text{s}}^{\text{local}}$ holds
only for values of $\Gamma$ which would appear to be unreasonably
small.  As already shown in Fig.~\ref{fig:2}, at low frequencies one
is more likely to find 
$R_{\text{s}}^{\text{nonlocal}} < R_{\text{s}}^{\text{local}}$.

The inset shows $R_{\text{s}}(\Gamma)$ at 220\,GHz. At this frequency
quasiparticles from the order parameter nodes can contribute to the
photon absorption and pair breaking yields a sizeable contribution to
$R_{\text{s}}$ so that nonlocal corrections are positive for all
$\Gamma$, diminishing as $\Gamma$ increases.

\noindent
\begin{minipage}{8.6cm}
  \centering%
  \begin{figure}[htb]
  \mbox{\psfig{file=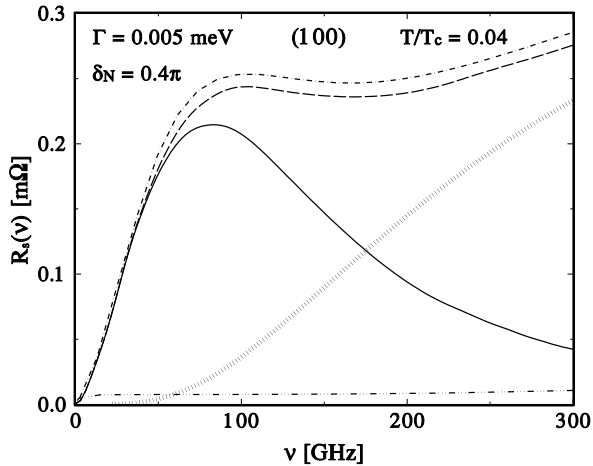,width=8.5cm}}
  \caption{\label{fig:5}
           Surface resistance as function of frequency. Comparison of
	   results including a small amount of scattering obtained
	   according to section~\ref{sec:theory,general} with results
	   obtained by taking the clean limit.
           Dot-dot-dashed line: local limit.
           Dot-dashed line: nonlocality included.
           Solid line: quasiparticle contribution in the clean limit
	   (\ref{eqn:tb12}).
           Dotted line: pair breaking contribution in the clean limit
	   (\ref{eqn:tb12}). 
           Dashed line: $R_{\text{s}}(\nu)$ in the clean limit.}
  \end{figure}
\end{minipage}

\bigskip

Fig.~\ref{fig:5} shows $R_{\text{s}}(\nu)$ for a very small scattering
rate, but for finite $\delta_{\text{N}}$. In this figure one can
clearly distinguish the quasiparticle contribution (solid line) and
the pair breaking contribution (dotted line) calculated from
eq.~(\ref{eqn:tb12}). The sum of these two contributions, shown as
dashed line, agrees well with the fully nonlocal calculations
according to section~\ref{sec:intro}.
These results are similar to those shown in Fig.~\ref{fig:1} and
demonstrate that an increase in $\delta_{\text{N}}$ can be compensated
by a reduction in $\Gamma$.

Finally, we turn to the other extreme geometry in which the order
parameter nodes are perpendicular and parallel to the surface
[(110)-surface, Fig.~\ref{fig:6}]. In this case, nonlocal effects 
always lead to an increase in $R_{\text{s}}$, but this is extremely
small even if a very small value of $\Gamma$ and $\delta_{\text{N}}=0$
is assumed. The corrections,
visible in our calculations only at low frequencies, represent the
contribution from quasiparticle states close to the order parameter
nodes. Momentum conserving pair breaking processes are negligible in
this geometry. This dramatic dependence of the nonlocal effects on the
geometric configuration is easily understandable in terms of the clean
limit formulae in section~\ref{sec:theory,clean}.

\noindent
\begin{minipage}{8.6cm}
  \centering%
  \begin{figure}[htb]
  \mbox{\psfig{file=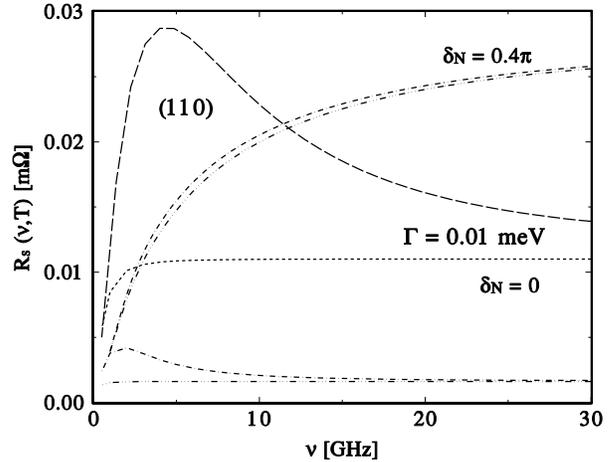,width=8.5cm}}
  \caption{\label{fig:6}
           Surface resistance as function of frequency for a different
	   orientation. of the order parameter relative to the surface
	   in the Born approximation for two different temperatures
	   (cf.\ Fig.~\ref{fig:1}). For $T/T_{\text{c}}=0.04$ we also
	   show results obtained by taking $\delta_{\text{N}}=0.4\,\pi$.}
  \end{figure}
\end{minipage}

\bigskip

\section{Conclusions}

It appears to be possible that nonlocal effects can be observed in the
surface resistance of high temperature superconductors, but only if
the order parameter has nodes. For an order parameter with
$d_{x^2-y^2}$-symmetry only (100) or (010) surfaces can be expected to
yield results significantly different  from the local
limit. Observation of such differences would provide direct evidence
for the order parameter symmetry. Two contributions to the nonlocal
response can be identified: absorption by quasiparticles and pair
breaking. The latter contribution dominates in the THz regime, but
since it is nearly frequency and temperature independent, it would be
hard to identify.

As in the anomalous skin effect in the normal state, the quasiparticle
contribution to $R_{\text{s}}$ goes through a maximum as the frequency
is increased when $\delta_{\text{N}}=0$. 
The peak height increases with the number of thermally
occupied quasiparticle states until the temperature is so high that
inelastic scattering severely limits the quasiparticle free mean paths. In
a search for this effect it is therefore, not advisable to perform the
experiments at very low temperatures. For the material parameter
chosen the increase in  $R_{\text{s}}$ over the local limit is largest
for frequencies near 100\,GHz. At the lower end of the microwave
regime, nonlocality actually reduces $R_{\text{s}}$ below the local
limit as a consequence of the anisotropy of the energy gap.

The size of nonlocal corrections to $R_{\text{s}}$ in the
superconducting state depends sensitively on the scattering phase shift
and not only on the disorder induced normal state scattering rate. If
disorder were correctly described by the strong scattering limit, for
nonlocal effects to be important a degree of perfection in the
CuO$_2$-planes would be required that appears to be unattainable. When
one moves away from this limit, the picture changes rapidly.

\acknowledgments
We are grateful to I.~Kosztin for providing us with his thesis and to
N.~Schopohl for helpful communications.
We would like to thank D. Fay for a careful reading of the manuscript.
This work has been funded by the {\em Deutsche Forschungsgemeinschaft}
through the {\em Graduiertenkolleg} ``Physik nanostrukturierter
Festk\"orper''.

\end{multicols}

\end{document}